\documentclass[twocolumn,aps,prl,preprintnumbers,showpacs,nofootinbib]{revtex4}
\usepackage[utf8]{inputenc}
\usepackage[english]{babel}
\usepackage{amsmath}
\usepackage{amsfonts}
\usepackage{amssymb}
\usepackage{epsfig}
\usepackage{graphics,psfrag,rotating}
\usepackage{graphicx}
\usepackage{dcolumn}
\usepackage{bm}
\usepackage{epstopdf}
\usepackage{color}
\usepackage[usenames,dvipsnames,svgnames]{xcolor}
\usepackage[colorlinks=true,
            linkcolor=red,
            urlcolor=gray,
            citecolor=blue]{hyperref}
  \usepackage{hyperref}

\newcommand{\lsim}   {\mathrel{\mathop{\kern 0pt \rlap
{\raise.2ex\hbox{$<$}}}
 \lower.9ex\hbox{\kern-.190em $\sim$}}}
\newcommand{\gsim}   {\mathrel{\mathop{\kern 0pt \rlap
{\raise.2ex\hbox{$>$}}}
\lower.9ex\hbox{\kern-.190em $\sim$}}}

\def\3nab{\tilde{\nabla}}

\def\hsp5{\hspace{5mm}}

\def\case#1/#2{\textstyle\frac{#1}{#2}}

\def\ber {\begin{eqnarray}}
\def\eer {\end{eqnarray}}
\def\bea {\begin{eqnarray}}
\def\eea {\end{eqnarray}}

\def\bc {\begin{center}}
\def\ec {\end{center}}
\def\case#1/#2{\frac{#1}{#2}}

\newcommand{\bw}{\begin{widetext}}
\newcommand{\ew}{\end{widetext}}

\newcommand{\be}{\begin{equation}}
\newcommand{\bse}{\begin{subequation}}
\newcommand{\ese}{\end{subequation}}
\newcommand{\ee}{\end{equation}}
\newcommand{\eei}{\end{eqnarray}\indent\indent}
\newcommand{\ba}{\begin{array}}
\newcommand{\ea}{\end{array}}
\newcommand{\bal}{\begin{eqnarray}}
\newcommand{\eal}{\end{eqnarray}}

\def\case#1/#2{\textstyle\frac{#1}{#2} }

\newcommand{\bver}{\begin{verbatim}}
\newcommand{\ever}{\end{verbatim}}

\begin{document}


\title{ Accelerating universe in Kaniadakis cosmology without need of dark energy}
\author{
A. Salehi\footnote{Email: salehi.a@lu.ac.ir},
}
\affiliation{ Department of Physics, Lorestan University, Khoramabad, Iran}

\date{\today}

\begin{abstract}
Taking into consideration of Kaniadakis entropy associated with the apparent horizon of Friedmann-Robertson-Walker
(FRW) Universe and using the gravity-thermodynamics conjecture, a new cosmological scenarios emerge based on corrected
Friedmann equations, which contains a correction term $ \alpha\left(H^2+\frac{k}{a^2}\right)^{-1}$ where $\alpha\equiv\frac{K^2 \pi^2}{2 G^2}$ and $K$ is  Kaniadakis parameter. We show that it is possible to reconstruct the parameters of the model, in terms of cosmographic parameters$\{ q, j, s\}$ analytically. For the flat universe, the parameters can be reconstructed in terms of only two cosmographic parameters $\{q, j\}$. The advantage of this analytical reconstruction is that not only it enables us to put constrain on parameters of the model in terms of directly measurable cosmographic parameters, but also provides the possibility to test observational measurements on Kaniadakis cosmology.
We examine the model for different sets of cosmographic parameters which observationally have been constraint in Kakidrosis cosmology. As an interesting result is that without any assumption about the value of $\Lambda$, we found that the set $\{q_{0}=-0.708, j_{0}=1.137\}$ automatically gives $\Lambda\simeq0$ and $\{\Omega_{m0}\simeq0.325,\Omega_{\alpha0}=0.671\}$. This result is in excellent agrement with pervious observational studies. Reconstructing  the evolution of deceleration parameter against redshift $z$ for these values, shows that the correction term could plays the role of dark energy without any dark energy component or cosmological constant $\Lambda$. Finally, we formulate the deviation parameter in terms of $\{q,j\}$ which reflects the deviation of  model from $\Lambda CDM$ model. We Show that the deviation factor is very sensitive to the jerk parameter $j$, while the $\Omega_{m0}$ is semiactive to deceleration parameter $q_{0}$. Hence the set ${j,q}$ can be regarded as useful parameters to test the theoretical and observational studies in Kaniadakis cosmology.
\end{abstract}


%
%


\maketitle
The quantum phenomenon of Hawking radiation \cite{Hawking}
indicates that black hole has a temperature proportional to
its surface gravity and an entropy proportional to its horizon
area \cite{Hawking,Bekenstein,Hooft,Susskind}. This issue led to an interesting exploration that expresses there is a profound
connection between gravity and thermodynamics which was
first addressed by Jacobson \cite{Jac}. This connection indicates that the field equations
of gravity can be written in the form of the first law of
thermodynamics on the horizon and vice versa (see
\cite{Jac,Pad1,Pad2,Pad3,CaiKim,Cai2,Shey1,Shey2,SheyCQ,SheyLog,SheyPL}
and references therein). It also indicates that
the field equations of gravity can be extracted from statistical
mechanics. In this approach, the gravitational field equations can be derived from the holographic principle and the equipartition
law of energy on the horizon degrees of freedom
 \cite{Ver}. This approach describes
  gravity as an entropic force
emerges from the change in the information of the system. A wide range of studies have focused on this scenario
 (see e.g.
\cite{Cai5,sheyECFE,ECEE,Visser,SRM}).

It is important to note that, in deriving the gravitational field equations
from the thermodynamic arguments, the entropy expression plays a
crucial role. Any modification to the entropy expression modifies
the corresponding field equations
\cite{CaiLM,SheT1,Odin1,SheT2,Emm2,SheB1,SheB2,Odin2,Odin3,Odin4,Odin5}.

The effects of the
generalized Kaniadakis entropy on the Friedmann equations have been investigated in \cite{Luci,Mor,Her,Dre,Kum1,Kum2}.
Modified Friedmann equations based on the generalized Kaniadakis
entropy was already investigated by two approaches. In the first approach which explored by \cite{Lym}, using the
relation $-dE=TdS $ on the apparent horizon of FRW universe, it was shown that the
Friedmann equations are modified by new extra terms that constitute an
effective dark energy sector depending on the Kaniadakis parameter
$K$.  Here
$-dE$ is the energy flux crossing the horizon within an
infinitesimal period of time $dt$, while $T$ and $S$ are,
respectively, the temperature and entropy associated with the
apparent horizon \cite{Lym}. In the second approach which recently presented by \cite{As},
 the geometry part of the cosmological field equations is modified,
and the energy content of the universe is not affected by the
generalized Kaniadakis entropy. While the author of
 \cite{As} stated that since entropy is a geometrical quantity and
any modification to it should change the geometry part
of the field equations, hence the second approach is more reasonable and. In addition, since our universe is
expanding thus we consider the work term (due to the
volume change) in the first law of thermodynamics and
write it as $dE= T dS+W dV$. In this paper, we
follow the second approach and use the modified Friedman equations obtained by\cite{As}, however, we show that the two approaches for $\alpha H^{-4}\equiv\frac{K^2 \pi^2}{2 G^2H^{4}}\ll1$, are equivalent. We reconstruct the parameters of the model  $\Omega_{m},\Omega_{\Lambda},\Omega_{\alpha}$ in terms of the cosmographic parameter $(q,j,s)$  which express geometrical property of the model. This approach provide us some advantages.

The structure of this paper is as follows. In section \ref{Kan},
we review the origin of the Kaniadakis entropy. In section \ref{FIRST}, we review the modified
Friedmann equations through the generalized Kaniadakis entropy from approaches explored by \cite{As} and \cite{Lym} and show that this two approaches are equivalent. In
section \ref{FIRST1}, we reconstruct the parameter of the model in terms of cosmographic parameters and constrain on parameters of the model . We finish with
conclusion in the last section.

\section{Kaniadakis horizon entropy \label{Kan}}
 Kaniadakis entropy is a
one-parameter entropy which generalizes the classical
Boltzmann-Gibbs-Shannon entropy. It originates from a coherent and
self-consistent relativistic statistical theory. The advantages of
Kaniadakis entropy is that it preserves the basic features of
standard statistical theory, and in the limiting case restore it
\cite{Kan1,Kan2}. The general expression of Kaniadakis entropy is
given by \cite{Kan1,Kan2}
\begin{eqnarray}
S_{K}=- k_{_B} \sum_i n_i\, \ln_{_{\{{\scriptstyle K}\}}}\!n_i  ,
\end{eqnarray}
with $k_{_B}$ is the Boltzmann constant, and
\begin{eqnarray}
\ln_{_{\{{\scriptstyle K}\}}}\!x=\frac{x^{K}-x^{-K}}{2K}.
\end{eqnarray}
Here $K$ is called the Kaniadakis parameter which is a
dimensionless parameter ranges as $-1<K<1$, and measures the
deviation from standard statistical mechanics. In the limiting
case where $K\rightarrow0$, the standard entropy is restored.

 Alternatively, Kaniadakis entropy
can be expressed as
\cite{Abreu:2016avj,Abreu:2017fhw,Abreu:2017hiy,Abreu:2018mti,Yang:2020ria,
Abreu:2021avp}
\begin{equation}
 \label{kstat}
S_{K} =-k_{_B}\sum^{W}_{i=1}\frac{P^{1+K}_{i}-P^{1-K}_{i}}{2K}.
\end{equation}
Here  $P_i$ is the probability in which the system to be in a
specific microstate and $W$ represents the total number of the
system configurations. Throughout this paper we set
$k_{_B}=c=\hbar=1$.

 apply the  Kaniadakis entropy in the
black hole thermodynamics gives

 \begin{equation} \label{kentropy}
S_{K} = \frac{1}{K}\sinh{(K S_{BH})}.
\end{equation}
Where
 \begin{equation} \label{kentropy2}
 S_{BH}=\frac{A}{4G}
\end{equation}
Where $S_{BH}$ is the standard Bekenstein-Hawking entropy, with $A$ as the horizon area and $G$ is is the gravitational constant. When $K\rightarrow 0$ one recovers the standard Bekenstein-Hawking
entropy, $S_{K\rightarrow 0}=S_{BH}$.
\section{Modified Friedmann equations from different approaches\label{FIRST}}
 Considering an expanding Universe filled with a matter
perfect fluid, with energy density $\rho_m$ and pressure $p_m$, which is
homogeneous and isotropic Friedmann-Robertson-Walker (FRW) geometry metric with form
\begin{equation}
ds^2=-dt^2+a^2(t)\left(\frac{dr^2}{1-kr^2}+r^2d\Omega^2 \right),
\label{metric}
\end{equation}
  where $a(t)$ is the scale factor, and with $k=0,+1,-1$ corresponding to flat,
close and open spatial geometry respectively. In order to exploit the gravitational thermodynamics conjecture in cosmology,
the
first law
is described in terms of the heat, considered as the energy that flows
through local Rindler horizons, applied on the horizon itself
\cite{Jac,Padmanabhan1,Padmanabhan2}, and in particular
on the apparent horizon
\cite{Bak:1999hd,Frolov:2002va,Cai:2005ra,Cai:2008gw}:
\begin{equation}
\label{apphor}
r_a=\frac{1}{\sqrt{H^2+\frac{k}{a^2}}},
\end{equation}
  where $H=\frac{\dot a}{a}$ the Hubble parameter and dots denoting derivatives
with respect to $t$.
The
authors of \cite{Lym} applied  the first law of thermodynamics as
\begin{equation}
-dE=TdS
\end{equation}
Here dE is the energy flux crossing the apparent
horizon within an infinitesimal period of time dt which can be expressed as
\begin{equation}\label{de0}
-dE=A(\rho_m+p_m)H r_{a}dt.
\end{equation}
Also they considered
Temperature as\begin{equation}
\label{Th}
 T=\frac{1}{2\pi r_a}.
\end{equation}

Then use the generalized Kaniadakis entropy, namely equation (\ref{kentropy}). In
particular, differentiating  (\ref{kentropy}) they derived
\be \label{dsk}
dS_{K}=\frac{8\pi}{4G}\cosh{\left(K  \frac{\pi r_a^2}{G}
\right)}r_{a}\dot{r_{a}}dt.
\ee
Applying the first
law of thermodynamics,
they derived the modified Friedman equations associated to Kaniadakis entropy as
\be \label{gfe1}
-4\pi G(\rho_{m}+p_{m})=\cosh{\left[K
\frac{\pi}{G(H^2+\frac{k}{a^2})}\right]}\left (\dot{H}-\frac{k}{a^2} \right ).
\ee

\begin{eqnarray} \label{gfe2}
&&
\!\!\!\!\!\!\!\!\!\!\!\!\!\!\!\!
\frac{8\pi G}{3}\rho_{m}= \cosh{\left[K
\frac{\pi}{G(H^2+\frac{k}{a^2})}\right]}\left (H^{2}+\frac{k}{a^2} \right
)
\nonumber\\
&&\ \ \ \ \
-\frac{K\pi}{G} \text{shi}{\left[K
\frac{\pi}{G(H^2+\frac{k}{a^2})}\right]}-\frac{\Lambda}{3},
\end{eqnarray}
 where $\Lambda$ is the integration constant and
The function $\text{shi}{(x)}$ is defined in
general as \begin{eqnarray}\text{shi}{(x)}=\int^{x}_{0}{\frac{\sinh(x')}{x'}dx'}\end{eqnarray}
Which is an entire
mathematical odd function of $x$ with no branch  discontinuities.

In the second approach which recently explored by \cite{As},
He take the first law of thermodynamics as
\begin{equation}\label{de2}
dE=TdS+WdV
\end{equation}
But here, $dE$ is the change
in the total energy inside the apparent horizon. Also $W$ is
work density associated with this volume change (due to
the universe expansion) and defined as\cite{Hayward}
\begin{equation}\label{de12}
W=\frac{1}{2}(\rho-P)
\end{equation}
The total energy of the
universe enclosed by the apparent horizon is $E = \rho V$, where $V=\frac{4}{3}\pi r_{a}^{3}$ is the volume enclosed by a 3-
dimensional sphere with the area of apparent horizon. Hence using conservation equation,
\begin{equation}
\label{dE2}
 dE=4\pi\tilde
 {r}_{A}^{2}\rho d\tilde {r}_{a}-4\pi H \tilde{r}_{a}^{3}(\rho_{m}+p_{m}) dt.
\end{equation}
Comparison the equations (\ref{dE2}) and (\ref{de0}) shows that. The difference between $dE$ in two approaches is $4\pi\tilde
 {r}_{A}^{2}\rho d\tilde {r}_{a}$. This difference due to this fact that the authors of \cite{Lym} assumed that within an infinitesimal period of time $dt$ the apparent horizon radius
$r_{a}$ is fixed. As another difference between two approach is that the associated temperature with the apparent horizon is
given by
\begin{equation}\label{T}
T=\frac{\kappa}{2\pi}=-\frac{1}{2 \pi \tilde
r_a}\left(1-\frac{\dot {\tilde r}_a}{2H\tilde r_a}\right).
\end{equation}
 Considering the fact that
deviation from the standard Bekenstein-Hawking is small, the author of \cite{As} assumed that
that $K\ll1$.  Hence the expression (\ref{kentropy}) can be expanded
as
\begin{equation}\label{kentropy2}
S_{K}=S_{BH}+ \frac{K^2}{6} S_{BH}^3+ {\cal{O}}(K^4).
\end{equation}
apply the above expression to
extract the modified friedmann equations, author of\cite{As} find the first and second modified Friedman equations as (For  more details see\cite{As})
\begin{equation} \label{Fried4}
H^2+\frac{k}{a^2}-\alpha\left(H^2+\frac{k}{a^2}\right)^{-1} =
\frac{8 \pi G}{3}(\rho_{m}+\rho_{\Lambda}).
\end{equation}
\begin{eqnarray}\label{2Fri2}
\left(\dot{H}- \frac{k}{a^2}\right)\left[1+\alpha
\left(H^2+\frac{k}{a^2}\right)^{-2}\right]=-4\pi G(\rho_{m}+p_{m}).
\end{eqnarray}
Here,
 \begin{eqnarray}
\alpha\equiv\frac{K^2 \pi^2}{2 G^2}
\end{eqnarray}
 When $\alpha\rightarrow 0$, the Friedmann
equations in standard cosmology are recovered.
\section{Equivalence of two approaches\label{FIRST2}}
Here we want to show that two approaches are not different and for, $KS_{BH}\ll1$, the condition which have been implemented in \cite{As}), the equations (\ref{gfe1}) and (\ref{gfe2}) are reduced to (\ref{Fried4}) and (\ref{2Fri2}). To prove this, we start from expanding  of $\cosh{x}$ and $\text{shi}{(x)}$
 \begin{align}
&\cosh{x}=1+\frac{x^{2}}{2!}+\frac{x^{4}}{4!}+..\\
&\text{shi}{(x)}=\int^{x}_{0}{\frac{\sinh(x')}{x'}dx'}=x+\frac{x^{3}}{3!3}+\frac{x^{5}}{5!5}+..
 \end{align}
 Hence for $x\ll1$, we can neglect the terms of orders  more than 2, hence we can use the  following approximations
  \begin{align}
&\cosh{x}\simeq1+\frac{x^{2}}{2!}\\
&\text{shi}{(x)}\simeq=x
 \end{align}
Since $K\ll1$ and consequently $K
\frac{\pi}{G(H^2+\frac{k}{a^2})}\ll1$, we can use the following approximations
  \begin{align}\label{cs}
&\cosh{\left[K
\frac{\pi}{G(H^2+\frac{k}{a^2})}\right]}\simeq1+(\frac{K^{2}\pi^{2}}{2G^{2}(H^2+\frac{k}{a^2})^{2}})\nonumber\\&=1+\alpha
\left(H^2+\frac{k}{a^2}\right)^{-2}\\
&\text{shi}{\left[K
\frac{\pi}{G(H^2+\frac{k}{a^2})}\right]}\simeq\frac{K\pi}{G(H^2+\frac{k}{a^2})}=\sqrt{2\alpha}\left(H^2+\frac{k}{a^2}\right)^{-1}\label{cs2}
 \end{align}
Inserting equations (\ref{cs}) and (\ref{cs2}) in equations (\ref{gfe1}) and (\ref{gfe2}), we arrive the equations (\ref{2Fri2}) and (\ref{Fried4}) respectively.
Hence, the apparent difference in the form of the equations is not related to the difference in the way the first law of thermodynamics is applied. Rather, in the second approach, approximation (\ref{kentropy2}) is used, and with this approximation, the first method is equivalent to second method

\section{Connection between parameters of the model and cosmographic parameters\label{FIRST1}}
In this section, we aim to reconstruct the parameters of the model in terms of cosmographic parameters.
Here, in order to simplify the equations of the model, we introduce following dimensionless variables
\begin{align}\label{new}
&\Omega_{m}=\frac{8\pi G\rho_{m}}{3H^{2}},\Omega_{\Lambda}=\frac{8\pi G \rho_{\Lambda}}{3H^{2}},\nonumber
  \\& \Omega_{k}=\frac{k}{a^{2}H^{2}}, \Omega_{\alpha}=\alpha\frac{\left(H^2+\frac{k}{a^2}\right)^{-1}}{H^{2}}
\end{align}
Hence using (\ref{new}), we can derive the following equations which arranged in a matrix  differential equation\\
$\frac{d}{dx}
\left(\begin{array}{c}
  \Omega_{m} \\
  \Omega_{k} \\
  \Omega_{\Lambda}\\H \end{array}\right) =\left(\begin{array}{cccc}
                                   -1+2q & 0 & 0&0 \\
                                   0 &  2q& 0&0 \\
                                   0 & 0 & 2+2q&0\\
                                   0 & 0 & 0&-1-q
                                 \end{array}\right)
  \left(\begin{array}{c}
  \Omega_{m} \\
  \Omega_{k} \\
  \Omega_{\Lambda}\\H \end{array}\right)$
  \begin {equation}\label{dif}
.
\end {equation}
Where, $x=\ln a$. By dividing both sides of the equations (\ref{Fried4}) and(\ref{2Fri2}) by $H^{2}$ and using relations (\ref{new}), the equations (\ref{Fried4}) and(\ref{2Fri2}) can be rewritten as,
 \begin{align}\label{new2}
 &1-\alpha H^{-4}\left(1+\Omega_{k}\right)^{-2}=\frac{\Omega_{m}+\Omega_{\Lambda}}{1+\Omega_{k}}\\
&\left(1+q+\Omega_{k}\right)\left[1+\alpha H^{-4}\left(1+\Omega_{k}\right)^{-2}\right]=\frac{3}{2}\Omega_{m}\label{new3}
\end{align}

Cosmography is the study of
scale factor by expanding it through the Taylor series with respect to the cosmic time. It is a powerful approach provides valuable
 information about the evolution of
the scale factor and its derivatives
This mathematical framework is inherently kinematic in the sense that it relies
only on geometrical assumptions for the metric and is
independent of cosmological model.\cite{Chiba,Caldwell,Visser,Dabrowski1,Dabrowski,Capozziello0,Cattoen,Maciej}.  To study cosmography, it is worthy of
introducing the cosmographic parameters as follows
define $n$th cosmographic parameter with $C^{(n)}$ as
 \begin{align}\label{cn0}
 C^{(n)}=\frac{1}{aH^{n}}\frac{d^{n}a}{dt^{n}}
 \end{align}
 Hence, it is easy to show that
\begin{align}\label{cn}
\frac{dC^{(n)}}{dx}=C^{(n+1)}+\Big((n-1)+nq\Big)C^{(n)}, \ \ n=0,1,2,....
\end{align}
Where,
\begin{align}
&C^{(0)}=H,\ \ C^{(1)}=-q,\ \ C^{(2)}=j, C^{(3)}=s,\ \ C^{(4)}=l,..
\end{align}
which are usually referred to as the Hubble, deceleration, jerk, snap and lerk parameters, respectively. So from equation (\ref{cn}), we can extract the following relations
\begin{align}
&\frac{dq}{dx}=-j+(1+2q)q,\label{qx}\\
&\frac{dj}{dx}=s+(2+3q)j\label{Qx}\\
&\frac{ds}{dx}=l+(3+4q)s\\
&\frac{dl}{dx}=m+(4+5q)l\\
&...\\
&\frac{dC^{(n)}}{dx}=C^{(n+1)}+\Big((n-1)+nq\Big)C^{(n)}
\end{align}

 From equations (\ref{new2}) and (\ref{new3}), one can obtain
 \begin{align}\label{new4}
 \frac{(\Omega_{m}+\Omega_{\Lambda})}{1+\Omega_{k}}+\frac{3\Omega_{m}}{2(1+q+\Omega_{k})}=2
\end{align}
Also the Kaniadakis parameter can be reconstructed as
 \begin{align}\label{new06}
 \alpha=\Big(1+\Omega_{k}-\Omega_{m}-\Omega_{\Lambda}\Big)(1+\Omega_{k})H^{4}
\end{align}
and
 \begin{align}\label{new060}
 \Omega_{\alpha}=\Big(1+\Omega_{k}-\Omega_{m}-\Omega_{\Lambda}\Big)
\end{align}
Taking differentiating of both sides of equation (\ref{new3}) with respect to $x=\ln a$, we arrive
 \begin{align}\label{new5}
&\left(\frac{dq}{dx}+\frac{d\Omega_{k}}{dx}\right)\left[1+\alpha H^{-4}\left(1+\Omega_{k}\right)^{-2}\right]\\ \nonumber
&+\left(1+q+\Omega_{k}\right)\Big(-4\alpha H^{-5}\frac{dH}{dx}\left(1+\Omega_{k}\right)^{-2}\\ \nonumber
&-2\alpha H^{-4}\left(1+\Omega_{k}\right)^{-3}\frac{d\Omega_{k}}{dx}\Big)=\frac{3}{2}\frac{d\Omega_{m}}{dx}
\end{align}
Hence, using equations (\ref{dif}),(\ref{new3}), equation (\ref{new5}) can be simplified as
 \begin{align}\label{new6}
\frac{dq}{dx}=&\frac{1}{3\Omega_m(1+\Omega_k)}\Big(8-21\Omega_mq-30\Omega_m\Omega_k+24q^2+24\Omega_k^2 \nonumber\\ &+24 q^2 \Omega_k+24 q \Omega_k^2-15 \Omega_m \Omega_k^2+48 q \Omega_k-21\Omega_m q \Omega_k \nonumber \\&+6\Omega_mq^2\Omega_k+8q^3+8\Omega_k^3+24\Omega_k+24q-15\Omega_m\nonumber\\
&-6\Omega_mq^2\Big)
\end{align}
Using equation (\ref{qx} and (\ref{new6}), we can drive
 \begin{align}\label{Q1}
j=&\frac{-1}{3\Omega_m(1+\Omega_k)}\Big(8-21\Omega_mq-30\Omega_m\Omega_k+24q^2+24\Omega_k^2 \nonumber\\ &+24 q^2 \Omega_k+24 q \Omega_k^2-15 \Omega_m \Omega_k^2+48 q \Omega_k-21\Omega_m q \Omega_k \nonumber \\&+6\Omega_mq^2\Omega_k+8q^3+8\Omega_k^3+24\Omega_k+24q-15\Omega_m\nonumber\\
&-6\Omega_mq^2\Big)+q+2q^{2}
\end{align}
Solving equation (\ref{Q1}) in terms of $\Omega_{m}$ gives
\begin{align}
&\Omega_{m}=\\
&\frac{8(1+3q^2+3\Omega_k^2+3q^2\Omega_k+3q\Omega_k^2+6q\Omega_k+q^3+\Omega_k^3+3\Omega_k+3q)}{3(8q+10\Omega_k+4q^2+5\Omega_k^2-j-j\Omega_k+8q\Omega_k+5)}\nonumber
\end{align}
Neglecting $\Omega_{k}$, it can be simplified in terms of $(q, j)$ as
\begin{align}\label{om}
\Omega_{m}=\frac{8}{3}\Big(\frac{1+3q+3q^2+q^3}{5-j+8q+4q^2}\Big)
\end{align}
Inserting this value in equation (\ref{new4}), parameter $\Omega_{\Lambda}$ is also obtained as
\begin{align}\label{lan}
\Omega_{\Lambda}=\frac{2}{3}\Big(\frac{4q^3+6q^2-5+3j}{-5+j-8q-4q^2}\Big)
\end{align}

Also from equation (\ref{new06}), parameter $\alpha$ can be obtained as
\begin{align}\label{alpha}
\alpha=\frac{j-1}{5-j+8q+4q^2}H^{4}
\end{align}
Consequently,
\begin{align}\label{oalpha}
\Omega_{\alpha}=\frac{j-1}{5-j+8q+4q^2}
\end{align}
Equations (\ref{om}), (\ref{lan}) and (\ref{alpha}) indicate that it is possible to reconstruct the parameters of the model, $(\Omega_{m},\Omega_{\Lambda},\Omega_{\alpha})$ in terms of directly measurable parameters $(q,j)$.  Also parameter $\alpha$ can be obtained in terms of $(H,q,j)$. Note that if $j \rightarrow 1$, then  $ \Omega_{\alpha}\rightarrow 0$ and viceversa. This result was predictable, because for a flat universe $k=0$, when $\Omega_{\alpha}\rightarrow 0$, the model reduces to the standard $\Lambda CDM$ model. Also by neglecting radiation density, the $\Lambda CDM$ model is described by $j=1$.
In fact for a flat $\Lambda CDM$ model according to the equation (\ref{new06}), $\Omega_{m}+\Omega_{\Lambda}=1$, so from equation (\ref{new4}) it is easy to get
\begin{align}\label{qq}
q=-1+\frac{3}{2}\Omega_{m}
\end{align}
Hence, $\frac{dq}{dx}=\frac{3}{2}\frac{d\Omega_{m}}{dx}$. So using relations (\ref{qx}) and (\ref{dif}), it will be obtained that $j=1$.
 (For more details see also \cite{Harrison}-\cite{Hut}) In principal the case $j=1$ is a third order ODE which is hold only for $\Lambda CDM$ model (for flat case $k=0 $ and neglecting radiation density energy $\Omega_{r}=0$). It indicates that based on the Friedman equation, the evolution of the Universe in $\Lambda CDM$ can be represented by the scale factor $a(t)$ in which $\frac{\dddot{a}}{aH^{3}}=1$.  Note that although the deceleration parameter changes in different epoches of the universe in $\Lambda CDM$ model, however the parameter $j$ remain unchanged in different apaches of the universe and it changes only when the model changes. So this parameter can be regarded as factor which enable us to distinguish between different cosmological models.
In principle the parameter $j$ as one of the pairs statefinder diagnostic $\{r\equiv j,s_{statefinder}\equiv\frac{j-1}{3(q-\frac{1}{2})}\}$ can distinguish $\Lambda CDM$ form other cosmological models(For more discussion one can see (\cite{Sahni2003}  and \cite{Alam}).
 Hence, it is expected that for those cosmological models whose behavior is more similar to the $\Lambda CDM$, the $j$ value be closer to one.
Therefore, it is expected that models whose behavior is more similar to a $\Lambda CDM$ such as  (CPL) model, the $\omega CDM$ and the $XCDM$ model will have their $j$ closer to 1.
There are some previous studies which have also reported $j\simeq1$  (i.e. the parameter $j$ should be near  $1$).
 There are also some previous studies which have obtained jerk parameter close to 1. For example see \cite{Maciej,Alam,Herrera-Zamorano,Escamilla,Capozz,Munoz,Escamilla2,Nikodem}. Hence, according to the equation (\ref{oalpha}) the parameter $\Omega_{\alpha}$ can also be regarded as a deviation factor which determine the amount of deviation from $\Lambda CDM$ model.
  We can rewrite equation (\ref{oalpha}) as
  \begin{align}\label{oalpha2}
\Omega_{\alpha}=\frac{j-1}{4(1+q)^{2}-(j-1)}
\end{align}
For $q_{0}\leq-0.5$, which is the result of many robust observations, then $4(1+q_{0})^{2}<1$, hence
  \begin{align}\label{oalpha3}
\frac{j-1}{4(1+q_{0})^{2}-(j_{0}-1)}>\frac{j_{0}-1}{1-(j_{0}-1)}
\end{align}
If we get $j_{d}=j_{0}-1$, then
 \begin{align}\label{oalpha3}
\frac{j_{d}}{1-j_{d}}<\Omega_{0\alpha}
\end{align}
Which yields
 \begin{align}\label{oalpha3}
j_{d}<\frac{\Omega_{0\alpha}}{1+\Omega_{0\alpha}},\ \ j_{0}<1+\frac{\Omega_{0\alpha}}{1+\Omega_{0\alpha}}
\end{align}
  Since, $0<\Omega_{0\alpha}<1$, so $(0<j_{d}<\frac{1}{2})$ and $(1<j_{0}<1.5)$. There are many independent studies that have investigated cosmographic parameters and put observational constraints on them. Here, we use the results of \cite{Her} which constrained on the cosmographic parameters $\{q_{0},j_{0}\}$ for two cases $\Lambda=0$ and $\Lambda\neq0$ using robust observations. For $\Lambda\neq0$ and $\Lambda=0$ the values $\{q_{0}=-0.708^{+0.016}_{-0.016}, j_{0}=1.137^{+0.014}_{-0.013}\}$ and $\{q_{0}=-0.610^{+0.028}_{-0.035},j_{0}=1.041^{+0.051}_{-0.036}\}$ was reported respectively.

At first we examined the case $\{q_{0}=-0.708^{+0.016}_{-0.016}, j_{0}=1.137^{+0.014}_{-0.013}\}$ which found for $\Lambda=0$. For this values, the equations (\ref{om}), (\ref{lan}) and (\ref{oalpha}) gives
 \begin{align}
\Omega_{m0}\simeq0.325, \Omega_{\Lambda0}\simeq0.003,\Omega_{\alpha0}\simeq0.671
 \end{align}
As an interesting result is that without making any assumption about the value of $\Lambda$, our analysis shows that for the case $\{q_{0}=-0.708, j_{0}=1.137\}$. $\Lambda\simeq0$. In order to have a more accurate comparison with the results of the \cite{Her}, we check the results obtained for the dimensionless parameter $\beta$, which has been defined as
\begin{align}
\beta\equiv\frac{K \pi}{GH_{0}^{2}}
 \end{align}
 Using equation (\ref{new}), we can express the parameter $\Omega_{0\alpha}$ in terms of this parameter as
 \begin{align}\label{ooo}
\Omega_{0\alpha}=\frac{\alpha}{H_{0}^{4}} =\frac{K^{2} \pi^{2}}{2 G^{2}H_{0}^{2}}=\frac{\beta^{2}}{2}
 \end{align}
The author of \cite{Her} for $\Lambda=0$ using joint sample found $\beta=1.161^{+0.013}_{-0.013}$, inserting this value in equation (\ref{ooo}) yields$\Omega_{0\alpha}=0.673$ which is in good agrement with that we have found as $\Omega_{0\alpha}=0.671$.
Also, for this case the author of \cite{Her} found $\Omega_{m0}\simeq0.325$ which is in good agrement with that we have found as$\Omega_{m0}\simeq0.325$.
 This results indicate that our analysis are consistence with that obtained by \cite{Her}, also confirm the validation of two studies, because from different approaches the same results have been obtained. In the next section, we explain the reason for the slight differences in the parameter values in our approach with \cite{Her}.
 Now we aim to check the results of \cite{Her} for case $\Lambda\neq0$. The cosmographic values which reported by \cite{Her} are as $\{q_{0}=-0.610^{+0.028}_{-0.035},j_{0}=1.041^{+0.051}_{-0.036}\}$.
 Inserting the above values in the equations (\ref{om}),(\ref{lan}) and (\ref{alpha}), the parameters of the model are obtained as
 \begin{align}
 \Omega_{m0}=0.278,\ \ \Omega_{\Lambda0}=0.651,\ \Omega_{\alpha0} =\simeq0.0703
 \end{align}
 Although the results of  $\Omega_{m0}=0.278$ and $\Omega_{\Lambda0}=0.651$ are comparable with many observation results, but the value of $\Omega_{\alpha0}$ is expected to be much smaller than $0.0703$.
  In order to carefully examine why the two models do not match very well in this case we recall the equation (\ref{oalpha3} and express it in terms of $\beta$ as
 \begin{align}\label{oalpha33}
j_{d}<\frac{\frac{\beta^{2}}{2}}{1+\frac{\beta^{2}}{2}}
\end{align}
Where for very small value of $\beta$, it can be simplified as $\frac{\beta^{2}}{2}$.
This equation indicates that for those values of $\beta$ that are of order $10^{-2}$, the value of $j_{d}$  must be approximately of order $10^{-4}$. For example if for the set $\{q_{0}=-0.610^{+0.028}_{-0.035},j_{0}=1.041^{+0.051}_{-0.036}\}$, we remain $q_{0}$ unchanged, but instead of $j_{0}=1.04$, we consider $j_{0}=1.001$ and $j_{0}=1.0001$,  according to the equations (\ref{om}),(\ref{lan}) and (\ref{alpha}),
The following results would be obtained
 \begin{align}
 &For\ \ j_{0}=1.001,\nonumber \\ &\ \ \Omega_{m0}=0.260,\ \ \Omega_{\Lambda0}=0.737,\ \Omega_{\alpha0} =\simeq 0.0016,\ \beta\simeq0.057\\
 &For\ \ j_{0}=1.0001,\nonumber \\ &\ \ \Omega_{m0}=0.260,\ \ \Omega_{\Lambda0}=0.739,\ \Omega_{\alpha0} =\simeq 0.0001,\ \beta\simeq0.018\label{ref2}
 \end{align}
  The authors of \cite{Her} for case $\Lambda\neq0$ and for different observational measurements (see Table 1) found that $\beta<0.01$,
According to our analysis (\ref{ref2}) in order to $\beta\simeq0.01$, ($j_{0}<1.0001$). In order to investigate the cause of this discrepancy and investigate whether this discrepancy is a serious problem or not we should note that the difference between what our theoretical model predicts $j_{0}=1.0001$ and what is observationally measured by \cite{Her} as $j_{0}=1.041^{+0.051}_{-0.036}$ is very small, also the value predicted by our approach is within $(1-\sigma)$ of the observational measurement, however this slight difference may be caused by the systematic errors or other errors . However this discussion has at least two interesting result as\\
The parameter $j_{0}$ in case$\Lambda\neq0$ is very close to 1 and a small change in the value of this parameter leads to a large change in the value of the parameters $\beta$ and $\Omega_{\alpha0}$. Hence parameter $j_{0}$ can be regarded as deviation parameter which measures the deviation of the model from $\Lambda CDM$ model.\\
2-Our theoretical approach could test the observational measurements in Kaniadakis cosmology and estimate their difference with the theory model.

Our analysis showed that $\Omega_{\alpha0}$ is very sensitive to changes of $j_{0}$, the question which may arises is that the parameter $\Omega_{m0}$ is sensitive to which of the parameters $q_{0}$ and $j_{0}$. In order to investigate this issue, we call the equation (\ref{om}). We can rewrite this equation in terms of $q_{0}$ and $j_{0}$ as
\begin{align}\label{om0}
\Omega_{m0}=\frac{8}{3}\Big(\frac{1+3q_{0}+3q_{0}^2+q_{0}^3}{4(1+q_{0})^{2}-(j_{0}-1)}\Big)
\end{align}
 For$\Lambda\neq0$, we showed that $j_{0}$ is very close to 1, hence $j_{0}-1\simeq0$ and we can neglect this term in comparison with $4(1+q_{0})^{2}$. So the equation (\ref{om0}) can be simplified as
\begin{align}\label{om0}
\Omega_{m0}\simeq\frac{8}{3}\Big(\frac{1+3q_{0}+3q_{0}^2+q_{0}^3}{4(1+q_{0})^{2}}\Big)=\frac{2}{3}(1+q_{0})
\end{align}
This is exactly the relation that hold for the $\Lambda CDM$. So we can conclude that changes of $\Omega_{m0}$ is sensitive to changes of $q_{0}$. Also, since $\Omega_{m0}+\Omega_{\alpha0}+\Omega_{\Lambda0}=1$, it can be concluded that the parameter $\Omega_{\Lambda0}$ is sensitive to both $q_{0}$ and $j_{0}$.

 The previous analysis was hold for $\Omega_{k}=0$. Here we aim to consider also $\Omega_{k}$ as well. for this case, we require the forth cosmographic parameter  $s$ to reconstruct  $(\Omega_{m},\Omega_{\Lambda},\Omega_{k},\alpha)$. Hence four unknown parameter  $(\Omega_{m},\Omega_{\Lambda},\Omega_{k},\alpha)$ can be reconstructed in terms of the four known parameter $(H,q,j,s)$. Hence, taking derivative in both sides of equation (\ref{Q1}) and using equation (\ref{Qx}), we can find $X$ as

\begin{align}\label{x0}
s=&-\frac{1}{6}(24q-24j+120q^3+96q^2+64q\Omega_{m}^2j+8\,{\Omega_k}^{2
}\nonumber\\&+120\Omega_{m}\Omega_{k}^3q^2+48q^4+15\Omega_{k}\Omega_{m}-24\Omega_{m}\Omega_{k}^2j\nonumber\\
&+96\Omega_{m}q^3\Omega_{k}+204\Omega_{m}q^2\Omega_{k}^2+54\Omega_{m}qj-80\Omega_{k}^4q^2\nonumber\\
&-12\Omega_{m}\Omega_{k}^3j+32q\Omega_{k}^3j-24\Omega_{m}q\Omega_{k}^2j
+12\Omega_{m}q^2\Omega_{k}\nonumber\\&+24q^2\Omega_{k}j+120\Omega_{m}q^3\Omega_{k}^2+15\Omega_{m}+16q\Omega_{k}^2-30\Omega_{m}q\nonumber\\
&+96q^2\Omega_{k}-160q^2\Omega_{k}^2-24q\Omega_{k}^3-108\Omega_{m}q^2-24q^3\Omega_{k}\nonumber\\
&-256q^3\Omega_{k}^2-224q^2\Omega_{k}^3-8q\Omega_{k}^4-24\Omega_{m}q\Omega_{k}j-24q^2j\nonumber\\
&+56q\Omega_{k}+30j\Omega_{m}-48qj-72\Omega_{m}q^3+24\Omega_{k}^3j\nonumber\\
&-24\Omega_{k}j-32q^4\Omega_{k}^2-160q^3\Omega_{k}^3+16\Omega_{k}^2j-60\Omega_{m}q\Omega_{k})\nonumber\\
&-64q^4\Omega_{k}
\end{align}
Hence, by solving simultaneously the equations (\ref{x0}), (\ref{Q1}), (\ref{new4}) and (\ref{new06}), we can reconstruct $(\Omega_{m},\Omega_{\Lambda},\Omega_{k},\alpha)$  in terms of the four known parameter $(H,q,j,s)$. For example if we use the previous values $q=-0.53, j=1.01$ and result of \cite{Capozziello} for $s$ which reported $s_{0}=-0.39$, we find
 \begin{align}
 \Omega_{m}=0.3172,\Omega_{\Lambda}=0.6735\ \ \Omega_{k}=0.00026,\ \ \alpha H_{0}^{-4}\simeq0.010
 \end{align}
Note that there are many independent bounds on the cosmographic parameters. Although they all obtained values for $H_{0}$ and $q_{0}$ that were not significantly different, they obtained different values for higher-order cosmographics $s_{0},l_{0},m_{0}$. Here, from the observational bounds reported for $s_{0}$ in the previous studies, we have selected $s_{0}=-0.39$ from\cite{Capozziello}. The reason for this selection is that this value is closer to the value that the $\Lambda CDM$ predicts theoretically. Since
by the same way that equation (\ref{qq}) was extracted, the cosmographic parameter $s$ in $\Lambda CDM$ model, can be obtained as
$
s=1-\frac{9}{2}\Omega_{m}
$ which implies that in the range $-\frac{2}{9}\simeq0.22<\Omega_{m}<\frac{4}{9}\simeq0.44$, the cosmographic parameter $s$ changes in the range [-1,0]. Hence from previous studies, we consider a case that bounded to this range.
\subsection{Analysis for case $\Lambda=0$}
In this subsection, we want to investigate the results in absence of cosmological constant $\Lambda$. For this case by considering a flat universe, we can benefit the equation (\ref{lan}). Hence by considering $\Omega_{\Lambda}=0$, the jerk parameter $j$ is simplified in terms of deceleration parameter $q$, as
   \begin{align}\label{qQ}
  j=-\frac{4}{3}q^3-2q^2+\frac{5}{3}
 \end{align}

 We can immediately test the results of \cite{Her} for $\Lambda=0$. As we discussed in the previous section  the best fitted cosmographic parameters $q_{0}, j_{0}\}$ in \cite{Her} was obtained as $\{q_{0}=-0.708, j_{0}=1.137\}$. Inserting the value $\{q_{0}=-0.708$ in equation (\ref{qQ}) gives $j_{0}=1.137331883$. This result not only confirms that both our analysis and analysis of \cite{Her} are correct, but also shows the sensitivity of the model parameters to changes in the $j$. As we discussed in previous section, for $j_{0}=1.137 $, we found $\Omega_{\Lambda}=0.003$. This means that a change of order $10^{-4}$ in $j$ leads to a change of order  $10^{-3}$ in $\Omega_{\Lambda}$ and consequently the other parameters of the model.
 Also by inserting $j$ from equation (\ref{qQ}) in to the equation (\ref{om}) and (\ref{alpha}), the matter energy density $\Omega_{m}$ and parameter $\alpha$ are obtained as
  \begin{align}\label{qm}
  &\Omega_{m}=\frac{4(1+q)}{2q+5}\\
 &\Omega_{\alpha}=-\frac{2q-1}{2q+5}\label{qm}
\end{align}
 Hence, we can reconstruct all of the parameters of the model in terms of deceleration parameter $q$. A wide range of observational studies on deceleration parameter, found $-1<q<-0.5$.
 Equation (\ref{om}) implies that for this range the matter energy density $\Omega_{m}$ changes from 0 to 0.5. and jerk parameter $j $ changes  from 1 to 1.33.

This is an interesting point which indicates that Kaniadakis cosmology could explain the current acceleration expansion of the universe without need of dark energy or cosmological constant $\Lambda$.
It is also interesting to note that for $\Lambda=0$, we can reconstruct the evolution of the universe only by knowing the current value of the deceleration parameter $q$. Since using the equations (\ref{qx}) and (\ref{qQ}),
  \begin{align}\label{qe}
  \frac{dq}{dx}=\frac{4}{3}q^3+4q^2+q-\frac{5}{3}
 \end{align}
 Which leads to following integral
  \begin{align}\label{qe2}
\int dx=3\int\frac{dq}{4q^3+12q^2+3q-5}
 \end{align}
 The solution of integral gives
  \begin{align}\label{qe3}
x=\frac{1}{6}\ln \Big (4-\frac{9}{(q+1)^{2}}\Big)+\frac{1}{6}\ln C
 \end{align}
 Where $ C$ is constant of integration.
 Since, $x=\ln a=-\ln(1+z)$, the equation (\ref{qe3})can be solved in terns of deceleration parameter $q$ as
 \begin{align}\label{qe4}
q=-1+\frac{3(1+z)^{3}}{\sqrt{4(1+z)^6-C}}
 \end{align}
By set $q_{0}$ for $z=0$ in equation(\ref{qe4}), we can reconstructed the constant $C$ in terms of$q_{0}$ as
\begin{align}
C=4-\frac{9}{(q_{0}+1)^{2}}
\end{align}
  By setting $q=0$ in equation (\ref{qe4}), the transition redshift can be obtained
 \begin{align}\label{zt}
z_{tr}=-1+\sqrt[6]{\frac{9}{5(q_{0}+1)^{2}}-\frac{4}{5}}
\end{align}
For example for $q_{0}=-0.708$, the equation (\ref{zt}) yields $z_{tr}\simeq0.6517$. This result is very close to results of (\cite{Her}) who found found $z_{tr}\simeq0.652^{+031}_{-0.031}$.
  The equation (\ref{qe4}) also shows that in absence of radiation energy density, for large value of $z$, the deceleration parameter $q\rightarrow\frac{1}{2}$.( when $ z\rightarrow\infty$ the second terms of r.h.s of the equation $\rightarrow\frac{3}{2}$).
  Note that equation (\ref{qe}) can also be solved numerically only by knowing $q_{0}$.
  For $q_{0}=-0.708$, we have plotted the evolution of deceleration parameter against redshift $z$. In top panel of Fig.1, the evolution of deceleration parameter against redshift $z$ has been show. Also the evolution of $\Omega_{m}$ and $\Omega_{\alpha}$ against redshift $z$ have been shown in low panel of Fig.1. As can be seen
the universe exhibits the usual thermal history, with the
sequence of matter and dark-energy eras with the transition from deceleration to acceleration  at $z_{tr}\simeq0.65$. Additionally, in the future the universe tends asymptotically to a complete dark-energy dominated, de-Sitter state.
   \begin{figure}[t]
\includegraphics[scale=.4]{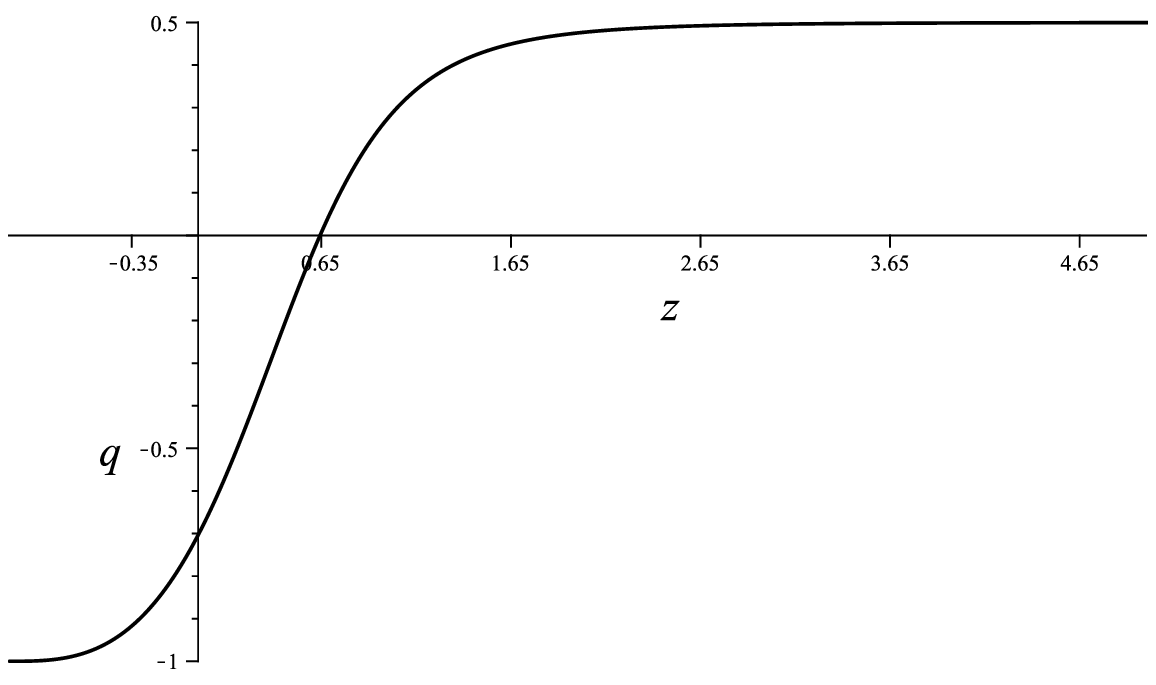}\hspace{0.1 cm}\\ \includegraphics[scale=.4]{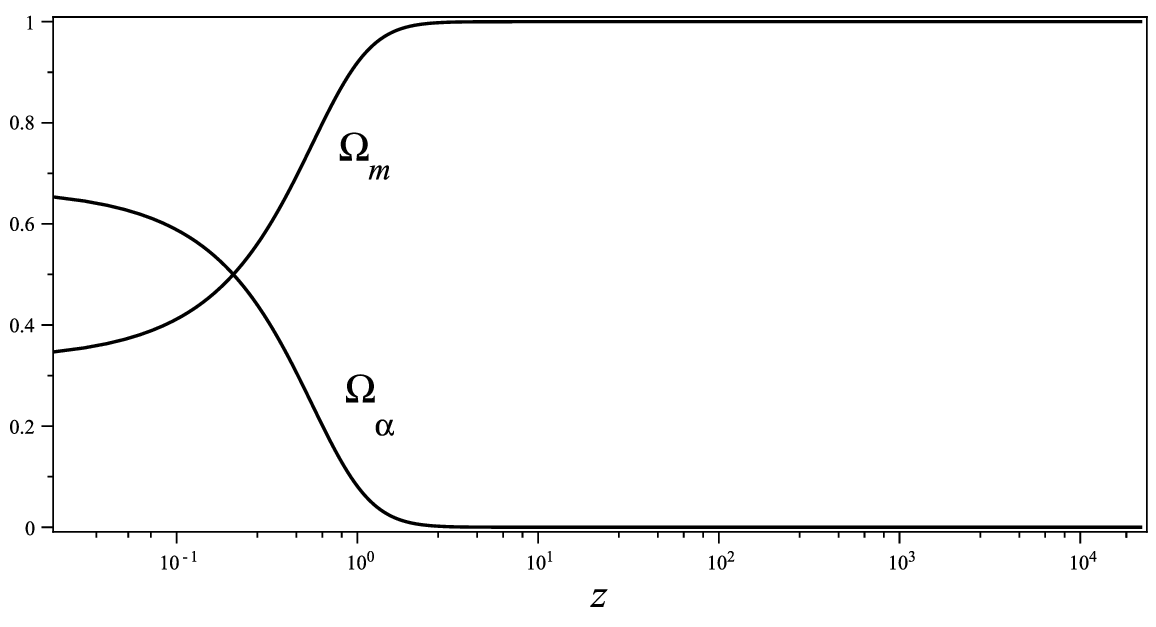}\hspace{0.1 cm}\\
Fig. 1: Top panel: the evolution of deceleration parameter against redshift $z$. Low panel: The evolution of $\Omega_{m}$ and $\Omega_{\alpha}$ against redshift $z$  \\
\end{figure}

 One can find easily from equations (\ref{new2}) and (\ref{new3}) that
  \begin{align}
  q=-\frac{4-5\Omega_{m}}{4-2\Omega_{m}}
 \end{align}

Also taking derivative respect to $x$ and using equation (\ref{qx}) and (\ref{dif}),
 \begin{align}
 j=\frac{64-240\Omega_{m}+412\Omega_{m}^2-309\Omega_{m}^3+82\Omega_{m}^4}{(2-\Omega_{m})^2(4-\Omega_{m})^2}
 \end{align}
By the same way, for $\Lambda=0$, we can find all of the cosmographic parameters in terms of $\Omega_{m}$.
 An interesting points that we found is that the modified term $\alpha\left(H^2+\frac{k}{a^2}\right)^{-2}$ could play a dual role, both the role of dark energy and the role of small correction in dark energy. For example for cosmographic set $q_{0}=-0.66,j_{0}=1.23,s_{0}=0.394$ which reported by \cite{Capozziello2}, our analysis gives
$
 \Omega_{m}=0.39,\Omega_{\Lambda}=0.01\ \ \Omega_{k}=0.024,\ \ \alpha H_{0}^{-4}\simeq0.60
$
\section{conclusion}
 Taking into account Kaniadakis entropy associated with the apparent horizon of Friedmann-Robertson-Walker
(FRW) Universe and using the gravity-thermodynamics conjecture, a new cosmological scenarios emerge based on corrected
Friedman equations. This modified equations in Kaniadakis entropy was first presented by \cite{Lym} then using different approach the author of \cite{As} find the modified equations which are apparently different from the equations of the first method.
At first we showed that this two approaches are equivalent. In other word, the modified Friedman equations derived by first approach for $K
\frac{\pi}{G(H^2+\frac{k}{a^2})}\ll1$ leads to those obtained in the second approach. In fact applying the first low of thermodynamic on Kaniadakis entropy, the modified Friedman equations contains a correction term $\alpha\left(H^2+\frac{k}{a^2}\right)^{-2} $ which could play different roles for different cases $\Lambda=0$ and $\Lambda\neq0$.

 In this paper we showed that it is possible to reconstruct parameters of the model $\{\Omega_{m},\Omega_{\Lambda},\Omega_{k}\}$ and $\Omega_{\alpha}$ or its equivalent $\beta$ in terms of directly measurable cosmographic parameters $\{q,j,s\}$. Establishing a relationship between the parameters that express energy content of the universe and cosmographic parameters that express the geometry of the universe  indicates how this two set of parameters affect by each other and is reminder of Willam's interesting interpretation of general relativity "matter tells space-time how to curve, and curved space-time tells matter how to move".\\
 Some of the interesting results of our analysis are as follows\\
 $\bullet$ Solving Friedman's equations analytically is only possible in very simple cases. But in this article, we related two set of parameters analytically, a set include energy component and another set include geometric parameters of the universe.
 If each set of the parameters be known the another set would be obtained obtained with high accuracy.\\

$\bullet$The cosmographic parameters are model independent parameters which are directly measurable. Hence, if we have enough quality observational data they can be constrained with high accuracy. Hence, in any cosmological model, if we could make a relation between parameters of the model with these parameters, we could make a good constrain on parameters of the model. Here we have made this relation and found the interesting results.\\

$\bullet$The relations, we extracted are analytical relation, in which, the model parameters are on one side  and cosmographic parameters on the other side of the equation. The efficiency of these equations is not limited to the estimation of the model parameters based on the cosmographic parameters, but these two way relations which enable us to test observational and numerical studies on Kaniadakis entropy and estimate the deviation of the observational measurement from that the analytical relation predicts.
 In order to test the efficiency of our model, we test previous study \cite{Her} which numerically and using robust observational have put constrain on both parameters of the model and cosmographic parameters. For case $\Lambda=0$, the author of \cite{Her} found $\{q_{0}=-0.708^{+0.016}_{-0.016}, j_{0}=1.137^{+0.014}_{-0.013}\}$.
 In order to test the validation of this results with our analytical approach, we insert these results in analytical equation without assumption that $\Lambda=0$, we found  $ \Omega_{\Lambda0}\simeq0.003,$ which is close to 0, also for this case we found $
\Omega_{m0}\simeq0.325,\Omega_{\alpha0}\simeq0.671$ which are very close to the results of \cite{Her}, who using joint sample found $\Omega_{m0}\simeq0.325$ and $\beta=1.161^{+0.013}_{-0.013}$ or $\Omega_{0\alpha}=0.673$. Interestingly we show that this slight difference is related to sensitivity of the model with parameter $j_{0}$. In other word our analytical model indicates that for this value  $\{q_{0}=-0.708, j_{0}=1.137331883\}$, the parameter $\Lambda$ will be exactly equal to zero. Hence the difference between $j_{0}=1.137331883$ and $j_{0}=1.137$ which is $\simeq0.0003$ leads to deviation value $\simeq0.003$ of parameter $ \Omega_{\Lambda0}$ from 0. Also for $q_{0}=-0.708$, the equation (\ref{zt}) yields $z_{tr}\simeq0.6517$. This result is very close to results of (\cite{Her}) who found found $z_{tr}\simeq0.652^{+031}_{-0.031}$. This results show that according to our analytical approach, the numerical analyzes in \cite{Her} have been done with very good accuracy so that they obtained a value for $j_{0}$ which has a very small deviation from that theoretical approach predicts.\\

$\bullet$We showed that for $\Lambda=0$, the correction term plays the role of dark energy and exhibits the thermal history of the universe which is expected evolution of the universe.\\

$\bullet$ As interesting result that theoretical approach predicts is that $\Omega_{\alpha0}$ or $\beta$ is very sensitive to the changes of $j_{0}$ and $\Omega_{m0}$ is very sensitive to changes of $q_{0}$. We also defined the deviation parameters $j_{d}$ which determine amount of deviation from $j_{0}=1$ which hold for ($\Lambda CDM$. and showed that for $q_{0}<-0.5$ the condition  $j_{d}<\frac{\beta^{2}}{2}$ is hold. Note that it is expected that the amount of deviation of the model from ($\Lambda CDM$ be more sensitive to changes of $j$, because in all of the history of evolution of the universe in $\Lambda CDM$ model $\Omega_{r}=0$, while the deceleration parameter changes from matter dominated with $q=\frac{1}{2}$ to dark energy dominated ($\Lambda$ dominated with $q=-1$), the jerk parameter is unchanged with $j=1$, hence $j$ changes only when the model deviate from $\Lambda CDM$. Therefore, it is expected that the greater the deviation of the model from the $\Lambda CDM$ model, the greater the difference of $j$ value with 1. \\

In summary, it can be concluded that as a useful approach, the theoretical approach could test the future theocratical and observational studies on kakidrosis entropy.

\section{Data availability statement}
The manuscript has no associated data or the data will not be deposited

\end{document}